\documentclass[10pt]{IEEEtran}
\usepackage{amsmath,amsfonts}
\usepackage{algorithmic}
\usepackage{array}
\usepackage{textcomp}
\usepackage{stfloats}
\usepackage{algorithm}
\usepackage{adjustbox}
\usepackage{cite}
\usepackage{graphicx}
\usepackage{amsmath,amssymb,amsfonts}
\usepackage{amsthm}
\usepackage{algorithmic}
\usepackage{graphicx}
\usepackage{textcomp}
\usepackage{xcolor}
\usepackage{algorithm}
\usepackage{array}
\usepackage{bm}
\usepackage{algorithm}
\usepackage{algorithmic}
\usepackage{caption}
\usepackage{subcaption}
\usepackage{url}
\usepackage{verbatim}
\usepackage{graphicx}
\usepackage{caption}
\usepackage{subcaption}
\usepackage{enumitem}
\usepackage[colorlinks]{hyperref}
\usepackage[dvipsnames]{xcolor}
\hypersetup{
    colorlinks = true,
    linkcolor  = blue,       
    citecolor  = blue,       
    urlcolor   = blue        
}
\def\BibTeX{{\rm B\kern-.05em{\sc i\kern-.025em b}\kern-.08em
    T\kern-.1667em\lower.7ex\hbox{E}\kern-.125emX}}
\usepackage{balance}
\begin{document}
\title{Safe World Model for Joint Trajectory and Beamforming Design of Mobile Base Stations}
\author{Abdolrasoul Sakhaei Gharagezlou, Zeinab Askari Donbeh, Mehdi Monemi, Matti Latva-aho, Mehdi Rasti
\thanks{All authors are with the Centre for Wireless Communications (CWC), University of Oulu, Finland (e-mail: \{abdolrasoul.sakhaeigharagezlou; zeinab.askaridonbeh; mehdi.monemi; matti.latva-aho; mehdi.rasti\}@oulu.fi).

}}

\IEEEaftertitletext{\vspace{-2.5 em}}
\maketitle

\begin{abstract}
Mobile base stations (MBSs), including aerial and ground-based platforms, provide flexible wireless coverage through adaptive positioning. However, jointly optimizing their trajectories and digital beamforming is challenging because movement changes channel conditions and interference, while beamforming determines the resulting signal and interference powers. These coupled decisions also account for how current movements affect future positions and communication performance under mobility and safety constraints. We formulate this problem as a constrained Markov decision process that maximizes the time-averaged sum signal-to-interference-plus-noise ratio (SINR), subject to mobility, obstacle region avoidance, and transmit-power constraints. To support long-term planning and reduce the extensive environment interactions required by model-free reinforcement learning, we develop a SafeDreamer-based world model framework that learns latent system dynamics from experience and optimizes a joint trajectory and beamforming policy through imagined trajectories. Feasibility-preserving action mappings and safety cost signals incorporate operational constraints into policy learning. Simulation results demonstrate that the proposed framework achieves higher time-averaged sum SINR than the proximal policy optimization and deep Q-network baselines, while exhibiting improved sample efficiency and more reliable constraint satisfaction.
\end{abstract}

\begin{IEEEkeywords}
Mobile base station (MBS), reinforcement learning (RL), world model, SafeDreamer.
\end{IEEEkeywords}
\vspace{-12pt}
\section{Introduction}
\IEEEPARstart{M}{OBILE} base stations (MBSs) enhance wireless coverage and communication performance through flexible deployment and adaptive mobility, particularly in areas with limited infrastructure or temporary communication demands \cite{uavIR, Ning}. However, jointly optimizing MBS trajectories and beamforming is challenging due to time-varying channels, mobility constraints, and temporal dependencies.

Extensive research has investigated MBS-assisted communications to improve security, coverage, and throughput \cite{uavIR, Ning, zhang, wu, sobhi, umar}. For example, the authors in \cite{uavIR} studied age-of-information minimization in unmanned aerial vehicle (UAV)-assisted covert communications in the presence of an aerial eavesdropper. The authors in \cite{Ning ,zhang} developed secure UAV-enabled integrated sensing and communication (ISAC) frameworks to maximize the secrecy rate. Despite these advances, conventional optimization methods often require accurate wireless environment models and incur high computational complexity. Such models are difficult to obtain under time-varying channel conditions and MBS mobility, while their high computational complexity limits real-time implementation. Although heuristic methods can alleviate these challenges, their scalability and robustness may remain limited in large-scale and dynamic wireless environments \cite{huristic}.

To overcome these limitations, recent studies have adopted reinforcement learning (RL) approaches, including model-based and model-free RL (MBRL/MFRL), to optimize system performance in dynamic wireless networks without requiring accurate mathematical models of the underlying network dynamics. For example, the authors in \cite{wu} proposed a hierarchical deep RL framework to enhance the performance of over-the-air federated learning systems in UAV-assisted wireless networks. The authors in \cite{sobhi} employed MBRL and MFRL for energy efficiency (EE) maximization in mmWave UAV networks, while the authors in \cite{umar} used proximal policy optimization (PPO) and deep deterministic policy gradient for sum-rate and EE maximization in multi-UAV networks. Nevertheless, existing RL-based approaches often suffer from {\it limited sample efficiency}, requiring extensive environment interactions to explore high-dimensional action spaces, while their {\it limited long-term planning capabilities} may hinder the optimization of temporally coupled trajectory and beamforming decisions.

World-model-based RL (WMBRL) offers a promising approach to addressing these challenges by learning compact representations of environment dynamics and optimizing policies through imagined trajectories. This enables sample-efficient policy optimization while accounting for the long-term effects of trajectory and beamforming decisions \cite{safedreamer, safedreamerV3}. {\it Such long-term planning is crucial for maximizing the time-averaged sum signal-to-interference-plus-noise ratio (SINR), as MBS mobility affects instantaneous communication performance, future channel conditions, and trajectory feasibility. However, jointly optimizing trajectory and beamforming through world-model-based long-term planning and safety-aware learning in dynamic MBS communications remains largely unexplored.}

Motivated by the above research gap, the main contributions of this paper are summarized as follows. First, we formulate the joint trajectory and digital beamforming optimization problem for MBSs as a constrained Markov decision process to maximize the time-averaged sum SINR under time-varying channel conditions, subject to MBS mobility, obstacle region (OR) avoidance, and transmit-power constraints, while accounting for the temporal dependencies of MBS trajectories. Second, we develop a WMBRL framework that learns network dynamics from collected experience and leverages imagined trajectories to jointly optimize MBS trajectories and digital beamforming, enabling long-term planning with fewer environment interactions. Third, we incorporate the constraints into policy learning through feasibility-preserving action mappings and safety-cost signals. Finally, simulation results demonstrate that the proposed framework achieves higher time-averaged sum SINR than the MFRL baselines, along with improved sample efficiency and more reliable constraint satisfaction.

\section{System Model and Problem Formulation}
We consider a communication system in which an MBS equipped with an \(N\)-element array serves \(K\) single-antenna users. The location of the \(k\)-th user is denoted by \(\mathbf{e}_k=[x_k,y_k,0]^T\), where \(x_k\) and \(y_k\) represent its horizontal coordinates. We consider an operation period of \(T^{\mathsf{tot}}\), divided into \(T^{\mathsf{s}}\) equal-duration time slots of length \(\delta=T^{\mathsf{tot}}/T^{\mathsf{s}}\) \cite{uavIR}. At time slot \(t\), the MBS moves horizontally at a fixed height \(L\), with its three-dimensional position denoted by \(\mathbf{e}^{\mathsf{MBS}}[t]=[x[t],y[t],L]^T\) \cite{uavIR}. Accordingly, the MBS's horizontal position at time slot \(t\) is expressed as \(\mathbf{u}[t]=[x[t],y[t]]^T\). The MBS displacement between consecutive time slots is constrained as
\begin{equation}\label{speed}
    \Vert {\mathbf{u}}[t+1]-\mathbf{u}[t]\Vert_2\le V^{\mathsf{max}}\delta, \quad t=0, \dots, T^{\mathsf{s}}-1,
\end{equation}
where $V^{\mathsf{max}}$ denotes the maximum movement speed of the MBS. Furthermore, the MBS moves from the initial position \(\mathbf u[0]=\mathbf u^{\mathsf I}\) to the final position \(\mathbf u[T^{\mathsf s}]=\mathbf u^{\mathsf F}\), where \(\mathbf u^{\mathsf I}=[x^{\mathsf I},y^{\mathsf I}]^T\) and \(\mathbf u^{\mathsf F}=[x^{\mathsf F},y^{\mathsf F}]^T\). Both the initial and final positions lie within the allowable operating region, satisfying \(x^{\mathsf{min}}\leq x^{\mathsf I},x^{\mathsf F}\leq x^{\mathsf{max}}\) and \(y^{\mathsf{min}}\leq y^{\mathsf I},y^{\mathsf F}\leq y^{\mathsf{max}}\), where \([x^{\mathsf{min}},x^{\mathsf{max}}]\) and \([y^{\mathsf{min}},y^{\mathsf{max}}]\) define the allowable coordinate ranges. In addition, the MBS must remain within the operating region, satisfying \(x^{\mathsf{min}}\leq x[t]\leq x^{\mathsf{max}}\) and \(y^{\mathsf{min}}\leq y[t]\leq y^{\mathsf{max}}\). The OR is modeled as a circular area that the MBS is required to avoid throughout its trajectory \cite{heo}. Let \(\mathbf u^{\mathsf{OR}}=[x^{\mathsf{OR}},y^{\mathsf{OR}}]^T\) and \(R^{\mathsf{OR}}\) denote the center coordinates and radius of the OR, respectively. Accordingly, the MBS trajectory must remain outside the OR throughout the operation period, expressed as follows \cite{heo}:
\begin{equation}\label{nfz}
\Vert\mathbf{u}[t]-\mathbf{u}^{\mathsf{OR}}\Vert_2
\geq R^{\mathsf{OR}}, 
\qquad t=0,\ldots, T^{\mathsf{s}}.
\end{equation}
To ensure that the MBS can reach its predefined final position \(\mathbf u^{\mathsf F}\), the initial and final positions are assumed to satisfy \(d^{\mathsf G}(\mathbf u^{\mathsf I},\mathbf u^{\mathsf F})\leq T^{\mathsf s}V^{\mathsf{max}}\delta\), where \(d^{\mathsf G}(\cdot,\cdot)\) denotes the shortest path length between two positions within the allowable operating region while avoiding the OR \cite{path}.
\vspace{-5pt}
\subsection{Time-Varying Channel and Signal Models}
The channel vector between the MBS and the \(k\)-th user at time slot \(t\) is expressed as \(\mathbf{g}_k[t]=\sqrt{\mu_k[t]}\widetilde{\mathbf{g}}_k[t]\), where \(\mu_k[t]=\frac{\mu_0}{D_k^2[t]}\) denotes the large-scale channel power gain, \(\mu_0\) represents the channel power gain at a reference distance of 1 m, \(D_k[t]=\|\mathbf{e}^{\mathsf{MBS}}[t]-\mathbf{e}_k\|_2\) denotes the MBS-user distance, and \(\widetilde{\mathbf{g}}_k[t]\in\mathbb{C}^{N\times 1}\) represents the corresponding small-scale fading vector. To capture temporal correlation across consecutive time slots, the small-scale fading follows a first-order complex Gauss-Markov process given by $\mathbf{\widetilde{g}}_k[t]=\rho \mathbf{\widetilde{g}}_k[t-1]+\sqrt{1-\rho^2}\,\boldsymbol{\epsilon}_k[t]$, where \(\widetilde{\mathbf{g}}_k[0]\sim\mathcal{CN}(\mathbf 0,\mathbf I_N)\), \(\boldsymbol{\epsilon}_k[t]\sim\mathcal{CN}(\mathbf 0,\mathbf I_N)\), and \(0\leq\rho<1\) denotes the temporal correlation coefficient of the small-scale fading between consecutive time slots \cite{channel}. The received signal at the $k$-th user in time slot $t$ is formulated as follows:
\begin{align}\label{signal}
    y_k\![t]\!= \mathbf{g}_k^H[t]\mathbf{w}_k[t]s_k[t]+\!\!\!\sum_{j=1,j\neq k}^{K}\!\!\! \mathbf{g}_k^H[t]\mathbf{w}_j[t]s_j[t] + n_k[t],
\end{align}
where \(\mathbf{w}_k[t]\) denotes the digital precoding vector for user \(k\) in time slot $t$, associated with the transmitted symbol \(s_k[t]\).  The transmitted symbols are normalized and mutually uncorrelated. Furthermore, $n_k[t] \sim \mathcal{CN}(0,\sigma_k^2)$ is the additive white Gaussian noise at user $k$ during time slot $t$. The SINR of the $k$-th user at time slot $t$ is formulated as $\Gamma_k[t] = \frac{\vert \mathbf{g}_k^H[t]\mathbf{w}_k[t]\vert^2}{\sum_{j=1,j\neq k}^{K} \vert \mathbf{g}_k^H[t]\mathbf{w}_j[t]\vert^2+\sigma^2_k}$.
\vspace{-5pt}
\subsection{Problem Formulation}
The objective of this paper is to maximize the time-averaged sum SINR of all users over a finite horizon of \(T^{\mathsf s}\) time slots. To achieve this objective, the MBS trajectory \(\{\mathbf{u}[t]\}\) and digital beamforming vectors \(\{\mathbf{w}_k[t]\}\) are jointly optimized under time-varying channel conditions, subject to initial and final position, mobility, operating-region, OR avoidance, and transmit-power constraints. The resulting optimization problem is formulated as follows:
\begin{subequations}\label{prob:SINR}
\begin{align}
\underset{
\{\mathbf{u}[t]\},
\{\mathbf{w}_k[t]\}
}{\operatorname{maximize}}
&\quad 
\frac{1}{T^\mathsf s}\sum_{t=1}^{T^{\mathsf s}}\sum_{k=1}^{K}\Gamma_k[t]
\label{prob:SINR_obj}
\\
\operatorname{s.t.}\quad
&
\left\{
\begin{aligned}
&\mathbf{u}[0]=\mathbf{u}^{\mathsf I},\quad
\mathbf{u}[T^{\mathsf s}]=\mathbf{u}^{\mathsf F},\\
&x^{\mathsf{min}}\leq x[t]\leq x^{\mathsf{max}},\quad t=0, \ldots, T^{\mathsf s},\\
&y^{\mathsf{min}}\leq y[t]\leq y^{\mathsf{max}},
\quad t=0, \ldots, T^{\mathsf s},\\
& \eqref{speed}, \eqref{nfz},
\end{aligned}
\right.
\label{prob:SINR_region}
\\
&
\sum_{k=1}^{K}\Vert\mathbf{w}_k[t]\Vert_2^2
\leq P^{\mathsf{max}},
\quad t=1, \ldots, T^{\mathsf s},  
\label{prob:SINR_power}
\end{align}
\end{subequations}
where $P^{\mathsf{max}}$ denotes the maximum transmit power at the MBS.

\section{Proposed Solution}
Problem \eqref{prob:SINR} is a non-convex stochastic optimization problem due to time-varying channels, coupled decision variables, and temporal dependencies in MBS trajectory optimization. Furthermore, the mobility, OR avoidance, and transmit-power constraints increase the complexity of the optimization. To address these challenges, we develop a SafeDreamer-based WMBRL framework that learns latent network dynamics from collected experience and jointly optimizes the MBS trajectory and digital beamforming through imagined trajectories, enabling long-term planning with fewer environment interactions. The framework handles the constraints through feasibility-preserving action mappings and safety-cost signals incorporated into policy learning. To apply the WMBRL framework to problem \eqref{prob:SINR}, we define the observation, action, and reward representations of the wireless environment as follows:
\vspace{-5pt}
\subsection{Observation, Action, and Reward Formulation}
\subsubsection{Observation Space}
At the beginning of time slot $t$, the observation vector is defined as follows:
\begin{equation}\label{statespace}
\begin{aligned}
\mathbf{o}[t] = \Big[
&{x}[t],\,{y}[t],\,
{D}_k[t], \Im\{\mathbf{g}_k[t]\}, \Re\{\mathbf{g}_k[t]\},\\
&
\Vert\mathbf{g}_k[t]\Vert^2_2,\,
{\Gamma}_k[t-1],\,
\mathbf{o}^{\mathsf{F}}[t],\,
\mathbf{o}^{\mathsf{OR}}[t]
\Big],
\end{aligned}
\end{equation}
where the complex channel vector $\mathbf{g}_k[t]$ is represented by its real and imaginary components, $\Im\{\mathbf{g}_k\}$ and $\Re\{\mathbf{g}_k\}$, respectively. Furthermore, \(\mathbf{o}^{\mathsf F}[t]=\big[x^{\mathsf F}-x[t],y^{\mathsf F}-y[t],B[t]-d^{\mathsf G}(\mathbf{u}[t],\mathbf{u}^{\mathsf F})\big]^T\) represents the final-position information, where \(B[t]=(T^{\mathsf s}-t)V^{\mathsf{max}}\delta\) denotes the maximum distance the MBS can travel within the remaining time slots. Similarly, \(\mathbf{o}^{\mathsf{OR}}[t]=\big[x[t]-x^{\mathsf{OR}},y[t]-y^{\mathsf{OR}},\|\mathbf{u}[t]-\mathbf{u}^{\mathsf{OR}}\|_2-R^{\mathsf{OR}}\big]^T\) represents the OR information, including the relative position of the MBS and its distance from the OR boundary. The observation-space dimension is \(d^{\mathsf o}=2NK+3K+8\).

\subsubsection{Action Space}
At each time slot \(t\), the action vector is defined as \(\mathbf{a}[t]=\big[\Delta \bar{x}[t],\,\Delta \bar{y}[t],\,\{\bar{P}_k[t]\}_{k=1}^{K}\big]^T\in[-1,1]^{d^\mathsf a}\), where \(\Delta\bar x[t]\) and \(\Delta\bar y[t]\) represent the MBS movement actions along the horizontal axes, while \(\bar P_k[t]\) denotes the power-control output for user \(k\). Furthermore, \(d^{\mathsf a}=K+2\) indicates the action-space dimension. The digital beamforming vector for user \(k\) is constructed as \(\mathbf{w}_k[t]=\sqrt{P_k[t]}\mathbf{q}_k[t]\), where \(P_k[t]\) is the allocated transmit power and \(\mathbf q_k[t]\) denotes the unit-norm regularized zero-forcing beamforming direction, obtained by normalizing the \(k\)-th column of $\mathbf{Q}[t] = \mathbf{G}[t](\mathbf{G}^H[t]\mathbf{G}[t]+\eta\mathbf{I}_K)^{-1}$, where $\mathbf{G}[t] = [\mathbf{g}_1[t], \dots, \mathbf{g}_K[t]]$ and  $\eta$ is the regularization parameter.

\subsubsection{Reward Function}
The immediate reward at time slot \(t\), designed to reflect the objective of problem \eqref{prob:SINR}, is defined as $r[t]=\sum_{k=1}^{K}\Gamma_k[t]$. The long-term return is defined as the cumulative discounted reward over the operation period $J^\mathsf r(\pi)=\mathbb{E}_{\pi}\left[\sum_{t=1}^{T^{\mathsf s}}\gamma^{t-1} r[t]\right]$, where \(\gamma\) denotes the discount factor and \(\pi\) represents the policy. Maximizing the discounted return encourages the agent to improve the long-term sum-SINR performance while accounting for the sequential effects of the MBS trajectory and digital-beamforming decisions over the operation period.
\vspace{-10pt}
\subsection{Constraint Handling}
The constraints in problem \eqref{prob:SINR} exhibit different characteristics and therefore require distinct constraint-handling approaches. Specifically, the MBS mobility and maximum transmit-power constraints are enforced through action-space construction. The initial-position constraint is satisfied by initializing the MBS at its predefined starting location, while the final-position constraint is enforced through trajectory projection to ensure that the MBS reaches its destination within the prescribed operation period.

\subsubsection{Action Level Constraints}
To satisfy the mobility constraint in \eqref{speed}, the movement action is mapped to a candidate MBS position \(\mathbf{u}^{\mathsf c}[t+1]\) as \(\mathbf{u}^{\mathsf c}[t+1]=\mathbf{u}[t]+V^{\mathsf{max}}\delta\frac{[\Delta\bar{x}[t],\Delta\bar{y}[t]]^T}{\max\{1,\|[\Delta\bar{x}[t],\Delta\bar{y}[t]]^T\|_2\}}\). This mapping ensures that \(\|\mathbf{u}^{\mathsf c}[t+1]-\mathbf{u}[t]\|_2\le V^{\mathsf{max}}\delta\), thereby satisfying the mobility constraint in \(\eqref{speed}\). The power-control outputs are mapped to feasible transmit-power allocations as \(P_k[t]=P^{\mathsf{max}}\frac{\exp(\bar{P}_k[t])}{\sum\_{j=1}^{K}\exp(\bar{P}_j[t])}\). This mapping guarantees that \(P_k[t]>0\) and \(\sum_{k=1}^{K}P_k[t]=P^{\mathsf{max}}\). Since \(\|\mathbf{q}_k[t]\|_2=1\), it follows that \(\sum_{k=1}^{K}\|\mathbf{w}_k[t]\|_2^2=\sum_{k=1}^{K}P_k[t]=P^{\mathsf{max}}\) \cite{RasoulWM}. Consequently, the transmit-power constraint in \eqref{prob:SINR_power} is satisfied.

\subsubsection{Feasibility-Preserving Trajectory Projection}
The initial-position constraint is satisfied by setting \(\mathbf u[0]=\mathbf u^{\mathsf I}\). To enforce the final-position constraint \(\mathbf u[T^{\mathsf s}]=\mathbf u^{\mathsf F}\), each candidate MBS position is evaluated for its reachability to the final destination within the remaining time slots while avoiding the OR. If the reachability condition is violated, the candidate position is projected onto a feasible position that satisfies the mobility and OR avoidance constraints while ensuring that the final destination remains reachable.

\subsubsection{Safety-Cost function} 
The safety-cost function comprises four components that quantify violations of the operating-region, final-position reachability, OR avoidance, and final-position constraints. Each component is evaluated using the candidate MBS position \(\mathbf u^{\mathsf c}[t+1]\) resulting from action \(\mathbf a[t]\), for \(t=0,\ldots,T^{\mathsf s}-1\), as detailed below.

\paragraph{Operating-region violation cost} This cost quantifies the deviation of the candidate MBS position from the allowable operating region and is defined as $c^{\mathsf{orv}}[t]=
\frac{\left\|\mathbf u^{\mathsf{c}}[t+1]-\Pi_{\mathcal R}\!\left(\mathbf u^{\mathsf{c}}[t+1]\right)\right\|_2}
{D^\mathsf{max}}$, where \(\Pi_{\mathcal R}(\cdot)\) represents the projection onto the allowable operating region and $D^\mathsf{max} = \sqrt{(x^{\mathsf{max}}-x^{\mathsf{min}})^2+(y^{\mathsf{max}}-y^{\mathsf{min}})^2}$. Accordingly, \(c^{\mathsf{orv}}[t]=0\) if the candidate MBS position lies within the allowable operating region. 

\paragraph{Final-position reachability cost} 
This cost quantifies the feasibility of reaching the final MBS position within the remaining operation period and is defined as $c^{\mathsf{fpv}}[t]=
\frac{\left[d^{\mathsf G}\!\left(\mathbf u^\mathsf c[t+1],\mathbf u^{\mathsf F}\right)-B[t+1]\right]^+}{D^\mathsf{max}}$,
where \([x]^+=\operatorname{max}\{x,0\}\). Accordingly, \(c^{\mathsf{fpv}}[t]=0\) when the MBS can reach its final position within the remaining mission duration.

\paragraph{OR violation cost}
This cost quantifies the violation of the OR avoidance constraint when the MBS enters the OR and is defined as \(c^{\mathsf{orc}}[t]=\frac{\left[R^{\mathsf{OR}}-\left\|\mathbf u^\mathsf c[t+1]-\mathbf u^{\mathsf{OR}}\right\|_2\right]^+}{R^{\mathsf{OR}}}\). Accordingly, \(c^{\mathsf{orc}}[t]=0\) when the MBS remains outside the OR.

\paragraph{Final-position error cost}
This cost quantifies the deviation of the MBS from its predefined final position at the end of the operation period and is defined as $c^{\mathsf{pev}}[t]
=\frac{\left\|\mathbf u^\mathsf c[t+1]-\mathbf u^{\mathsf F}\right\|_2}{D^\mathsf{max}}\mathbb{I}\!\left\{t=T^{\mathsf s}-1\right\}$, where $\mathbb{I}\{\cdot\}$ denotes the indicator function. Accordingly, this cost is evaluated only at the end of the operation period to assess whether the MBS reaches its predefined final position. Finally, the total safety cost incurred by action \(\mathbf a[t]\) at time slot $t$ is defined as $
c[t]=\frac{c^{\mathsf{orv}}[t]+c^{\mathsf{fpv}}[t]+c^{\mathsf{orc}}[t]}{T^\mathsf s}+c^{\mathsf{pev}}[t]$. The expected discounted safety return is defined as $J^{\mathsf c}(\pi)=\mathbb E_{\pi}\left[\sum_{t=0}^{T^{\mathsf s}-1}\gamma^{t}c[t]\right]$. Accordingly, the constrained policy optimization problem is formulated as follows:
\begin{equation}\label{LearningProb}
    \begin{aligned}
        \underset{\pi}{\operatorname{maximize}} \quad & J^\mathsf r(\pi) \quad \operatorname{s.t.}\,\, J^\mathsf c(\pi) \leq b,
    \end{aligned}
\end{equation}
where \(b\) denotes the predefined safety-cost budget. This formulation aims to maximize the long-term sum SINR while ensuring that the expected safety cost does not exceed the predefined budget \(b\). To solve the problem in \eqref{LearningProb}, the proposed learning framework is described as follows.
\vspace{-5pt}
\subsection{Learning Method}
The proposed WMBRL framework consists of three main components: (i) a \textbf{world model} that learns a compact latent representation from high-dimensional observations and predicts future states, rewards, and safety costs; (ii) \textbf{reward and cost critics} that estimate the expected long-term sum SINR and associated safety costs under a given policy; (iii) a \textbf{safe actor} that selects actions to maximize the expected long-term reward while satisfying the prescribed safety constraints. These components are detailed in the following subsections.

\subsubsection{World Model Learning}
The world model comprises an observation encoder, a recurrent state-space model (RSSM), and prediction heads for the observations, rewards, and safety cost signals. The observation encoder maps the high-dimensional observation \(\mathbf{o}[t]\) into a compact latent representation, while the RSSM models its temporal evolution based on the agent's action \(\mathbf{a}[t]\). Specifically,  at each time slot $t$, the deterministic state \(\mathbf{h}[t]\) is updated based on the previous deterministic state \(\mathbf{h}[t-1]\), stochastic state \(\mathbf{z}[t-1]\), and action \(\mathbf{a}[t-1]\), while the stochastic state \(\mathbf{z}[t]\) is inferred from the current deterministic state \(\mathbf{h}[t]\) and observation \(\mathbf{o}[t]\). The RSSM models the environment dynamics through its sequence model and transition prior, enabling the generation of imagined trajectories without direct environment interactions. The resulting latent states are used to reconstruct observations and predict rewards, safety-cost, and continuation signals. Accordingly, the components of the proposed world model are formulated as follows \cite{safedreamer}:
\begin{equation*}
\begin{aligned}
\mathrm{RSSM}
&\left\{
\begin{aligned}
&\text{Sequence model:}\,
\mathbf{h}[t]
\!=\!\!f_{\phi}\!\left(
\mathbf{h}[t\!-\!1],\mathbf{z}[t\!-\!1],\mathbf{a}[t\!-\!1]
\right),\\
&\text{Encoder:}\,
\mathbf{z}[t]
\sim q_{\phi}\!\left(
\mathbf{z}[t]\mid \mathbf{h}[t],\mathbf{o}[t]
\right),\\
&\text{Transition prior:}\quad
\hat{\mathbf{z}}[t]
\sim p_{\phi}\!\left(
\hat{\mathbf{z}}[t]\mid \mathbf{h}[t]
\right),
\end{aligned}
\right.
\\[1mm]
&\text{Reward predictor:}\,
\hat{r}[t]\sim p_{\phi}\!\left(
\hat{r}[t]\mid \mathbf{h}[t],\mathbf{z}[t]
\right),\\
&\text{Cost predictor:}\,
\hat{c}[t]\sim p_{\phi}\!\left(
\hat{c}[t]\mid \mathbf{h}[t],\mathbf{z}[t]
\right),\\
& \text{Continuation predictor:} \hat{d}[t]\sim p_\phi(\hat{d}[t] \mid  \mathbf{h}[t],\mathbf{z}[t]), \\
&\text{Decoder:}\quad
\hat{\mathbf{o}}[t]\sim p_{\phi}\!\left(
\hat{\mathbf{o}}[t]\mid \mathbf{h}[t],\mathbf{z}[t]
\right),
\end{aligned}
\end{equation*}
where \(\phi\) denotes the combined learnable parameters of the world model, while \(\hat{\mathbf z}[t]\), \(\hat r[t]\), \(\hat c[t]\), \(\hat d[t]\), and \(\hat{\mathbf o}[t]\) represent the predicted latent state, reward, safety cost, continuation, and reconstructed observation, respectively. The functions \(q_\phi(\cdot)\) and \(p_\phi(\cdot)\) denote the posterior and predictive distributions, while \(f_\phi(\cdot)\) represents the deterministic state-transition function. The total world-model loss is formulated as follows \cite{safedreamerV3}:
\begin{equation}\label{total-loss}
\mathcal{L}(\!\phi\!)\!\!=\!\mathbb{E}_{q_{\phi}}\!\!\left[\!\sum_{t=1}^{T^\mathsf s}\!\!\left(\beta^{\mathsf{pred}} \mathcal{L}^{\mathsf{pred}}[t]\!+\!\beta^{\mathsf{tran}} \mathcal{L}^{\mathsf{tran}}[t]\!+\!\beta^{\mathsf{rep}}\mathcal{L}^{\mathsf{rep}}[t]\right)\!\!\right]\!,
\end{equation}
where \(\mathcal{L}^{\mathsf{pred}}\), \(\mathcal{L}^{\mathsf{tran}}\), and \(\mathcal{L}^{\mathsf{rep}}\) denote the prediction,  transition, and representation losses, respectively, with corresponding weighting coefficients \(\beta^{\mathsf{pred}}=1\), \(\beta^{\mathsf{tran}}=0.5\), and \(\beta^{\mathsf{rep}}=0.1\). The prediction loss jointly trains the observation decoder, reward, safety cost, and continuation predictors as follows:
\begin{equation*}
\begin{aligned}
\mathcal{L}^{\mathsf{pred}}[t]
=&-\log p_{\phi}\!\left(
\hat{\mathbf{o}}[t]\mid\mathbf{h}[t],\mathbf{z}[t]
\right) -\log p_{\phi}\!\left(
\hat{r}[t]\mid\mathbf{h}[t],\mathbf{z}[t]
\right) \\
&-\log p_{\phi}\!\left(
\hat{c}[t]\mid\mathbf{h}[t],\mathbf{z}[t]
\right)-\log p_{\phi}\!\left(
\hat{d}[t]\mid\mathbf{h}[t],\mathbf{z}[t]
\right).
\end{aligned}
\label{eq:prediction_loss}
\end{equation*}
The transition loss minimizes the Kullback–Leibler (KL) divergence between the predicted prior and posterior latent distributions and is defined as follows \cite{safedreamerV3}:
\begin{equation*}
\mathcal{L}^{\mathsf{tran}}[t]\!=\!\operatorname{max}\!\left\{\!
\!1,\operatorname{KL} \!\left[\operatorname{sg}\!\left(q_{\phi}\!\left(\mathbf{z}[t]\!\mid\!\mathbf{h}[t],\mathbf{o}[t]\right)\!\right)\!\Big\|p_{\phi}\!\left(\mathbf{z}[t]\!\mid\!\mathbf{h}[t]\right)\right]\!\right\},
\label{eq:dynamics_loss}
\end{equation*}
where $\operatorname{sg}(\cdot)$ denotes the stop-gradient operator. The representation loss minimizes the KL divergence between the posterior and predicted prior latent distributions to train the posterior encoder and is defined as follows \cite{safedreamerV3}:
\begin{equation*}
\mathcal{L}^{\mathsf{rep}}[t]\!\!=\!\max\!\left\{\!1,{\operatorname{KL}}\!\left[q_{\phi}\!\left(\mathbf{z}[t]\!\!\mid\!\!\mathbf{h}[t],\mathbf{o}[t]\right)\!\Big\|\!\operatorname{sg}\!\left(p_{\phi}\!\left(\mathbf{z}[t]\!\mid\!\mathbf{h}[t]\right)\right)\right]\!\right\}.
\label{eq:representation_loss}
\end{equation*}
These losses enable the world model to learn compact latent representations and predict future states, rewards, and safety costs for policy optimization over imagined trajectories.
\subsubsection{Critic Learning}
The proposed framework utilizes two critic networks to estimate the expected long-term sum SINR and safety costs of the learned policy based on imagined trajectories generated by the world model. The reward and cost critics estimate the long-term cumulative reward and safety cost over these trajectories, respectively. The critics are trained using temporal-difference TD(\(\lambda\)) returns, which combine predicted immediate rewards or costs with their corresponding future value estimates.
Accordingly, the reward TD(\(\lambda\)) return is recursively defined as follows\cite{safedreamerV3}:
\begin{equation*}\label{eq:reward_td}
V_{\lambda}^{\mathsf r}[t]\!\!=\!\!
\begin{cases}
\!\hat r[t]\!\!+\!\!\nu[t]\!\left[\!
(1\!\!-\!\lambda)v_{\psi^{\mathsf r}}\!
\left(\mathbf h[t\!+\!1],\mathbf z[t\!+\!1]\right)\!
\!+\!\!\lambda V_{\lambda}^{\mathsf r}[t\!+\!1]
\right]\!, &\!\!\!\! \!\!t\!<\!\!\widetilde{H},\\[2mm]
v_{\psi^{\mathsf r}}
\left(\mathbf h[\widetilde{H}],\mathbf z[\widetilde{H}]\right), &\!\!\!\!\!\!\!\! t\!=\!\widetilde{H},
\end{cases}
\end{equation*}
where $\nu[t] = \gamma\hat{d}[t]$, $v_{\psi^{\mathsf r}}(\cdot)$ denotes the reward critic with trainable parameters $\psi^{\mathsf r}$, $\widetilde{H}$ represents the imagination horizon, and $\lambda$ is the TD($\lambda$) parameter.  Similarly, the cost TD(\(\lambda\)) return is formulated as follows:
\begin{equation*}\label{eq:cost_td}
V_{\lambda}^{\mathsf c}[t]\!=\!\!
\begin{cases}
\!\!\hat c[t]\!+\!\nu[t]\!\left[\!
(\!1\!\!-\!\lambda)v_{\psi^{\mathsf c}}
\!\left(\mathbf h[t\!+\!1],\mathbf z[t\!+\!1]\right)\!\!
+\!\!\lambda V_{\lambda}^{\mathsf c}[t\!+\!1]
\right]\!, & \!\!\!\!\!\!t\!<\!\!\widetilde{H}\!,\\[2mm]
v_{\psi^{\mathsf c}}
\left(\mathbf h[\widetilde{H}],\mathbf z[\widetilde{H}]\right), & \!\!\!\!\!\!\!\!t\!=\!\!\widetilde{H}\!,
\end{cases}
\end{equation*}
where \(v_{\psi^{\mathsf c}}(\cdot)\) denotes the cost critic, parameterized by the trainable parameters \(\psi^{\mathsf c}\). The reward and cost critics estimate $V^\mathsf r_{\lambda}[t]$ and $V^\mathsf c_{\lambda}[t]$, respectively, using Dreamer-style two-hot categorical targets in the symlog domain. Accordingly, the reward-critic loss is given by $\mathcal{L}(\psi^\mathsf r)=\mathbb{E}_t\left[-\operatorname{twohot}\!\left(\operatorname{symlog}\!\left(V_{\lambda}^{\mathsf r}[t]\right)\right)^{T}\log {p}_{\psi^\mathsf r}[t]\right]$,
while the cost-critic loss is given by $\mathcal{L}(\psi^\mathsf c)=\mathbb{E}_t\left[-\operatorname{twohot}\!\left(\operatorname{symlog}\!\left(V_{\lambda}^{\mathsf c}[t]\right)\right)^{T}\log {p}_{\psi^\mathsf c}[t]\right]$, in which $\operatorname{symlog}(x)=\operatorname{sign}(x)\log\!\left(|x|+1\right)$ \cite{safedreamerV3}. The reward critic estimates the expected long-term sum SINR, while the cost critic estimates the cumulative safety costs.

\subsubsection{Safe Actor Learning}
The actor is optimized to maximize the time-averaged sum SINR while reducing safety-constraint violations by leveraging imagined trajectories generated by the world model and future return estimates provided by the reward and cost critics. To this end, the reward and cost advantages are computed using the corresponding TD(\(\lambda\)) returns and critic estimates as \(A^{\mathsf r}[t]=\frac{V_{\lambda}^{\mathsf r}[t]-v_{\psi^{\mathsf r}}[t]}{S^{\mathsf r}}\) and \(A^{\mathsf c}[t]=\frac{V_{\lambda}^{\mathsf c}[t]-v_{\psi^{\mathsf c}}[t]}{S^{\mathsf c}}\), where \(S^{\mathsf r}\) and \(S^{\mathsf c}\) denote the reward and cost return-normalization scales, respectively. The reward and cost advantages are combined using the Lagrange multiplier \(\lambda^{\mathsf p}\) as $A[t]=\frac{A^\mathsf r[t]-\lambda^\mathsf p A^\mathsf c[t]}{1+\lambda^\mathsf p}$. The Lagrange multiplier $\lambda^\mathsf p$ is updated using a proportional-integral-derivative (PID) mechanism based on the observed safety-cost return, expressed as \(\lambda^{\mathsf p}=[\alpha(J^{\mathsf c}- b)+I+\tau\zeta]_{0}^{\lambda^{\mathsf{max}}}\), where \(\alpha\) and \(\tau\) denote the proportional and derivative gains, respectively, \(I\) represents the integral term, \(\zeta=\max\{0,J^{\mathsf c}-\bar{J}^{\mathsf c}\}\) captures increases in safety-cost returns between consecutive training iterations, and \(\bar{J}^{\mathsf c}\) denotes the safety-cost return from the previous iteration. Furthermore, the integral term \(I\) is updated as \(I\leftarrow[I+\kappa(J^{\mathsf c}-b)]_{0}^{\lambda^{\mathsf{max}}}\), where \(\kappa\) denotes the integral gain, \(\lambda^{\mathsf{max}}\) represents the maximum allowable Lagrange multiplier, and \([x]_{0}^{\lambda^{\mathsf{max}}}=\min\{\lambda^{\mathsf{max}},\max\{0,x\}\}\) denotes the clipping operation \cite{PID}. Accordingly, the actor loss is defined as 
\begin{align}
    \mathcal{L}(\boldsymbol{\theta}) =-\mathbb{E}_{t}\Big[
\Big(\mathrm{sg}\big(A[t]\big)&\log \pi_{\boldsymbol{\theta}}
\big(\mathbf{a}[t]\mid\mathbf{h}[t],\mathbf{z}[t]\big)\nonumber\\&+\xi\mathcal{H}\big[
\pi_{\boldsymbol{\theta}}(\cdot\mid\mathbf{h}[t],\mathbf{z}[t])\big]\Big)\Big],
\end{align}
where \(\pi_\theta(\cdot)\) denotes the stochastic actor policy parameterized by \(\boldsymbol{\theta}\), \(\mathcal H[\pi_\theta]\) is its entropy, and \(\xi\) is the entropy regularization coefficient. This mechanism enables safety-aware long-term reward maximization through imagined trajectories.
\section{Simulation Results}
\vspace{-5 pt}
We consider an \(N=10\)-antenna MBS serving \(K=2\) users over \(T^{\mathsf s}=100\) time slots, each lasting \(\delta=0.5\) s. The MBS operates at a fixed height of \(L=10\) m within a \(500\times500~\mathrm{m}^2\) region, traveling from \(\mathbf u^{\mathsf I}=(40,40)\) m to \(\mathbf u^{\mathsf F}=(460,460)\) m at a maximum speed of \(V^{\mathsf{max}}=25\) m/s. A circular OR with radius \(R^{\mathsf{OR}}=120\) m is centered at \((350,250)\) m. Unless otherwise stated, \(P^{\mathsf{max}}\!=\!\!1\) W, \(\sigma_k^2\!=\!-90\) dBm, \(\rho\!=\!0.9\), and $\mu_0\!=\!-30$ dB. The RSSM uses a 256-dimensional deterministic state and a stochastic state of 16 categorical variables with 16
classes each. We set the imagination horizon \(\widetilde{H}\!=\!\!12\), discount factor \(\gamma\!=\!0.997\), TD(\(\lambda\)) parameter \(\lambda=0.95\), and batch size \(16\). The world-model and actor/critic learning rates are \(10^{-4}\) and \(3\times10^{-5}\), respectively. The critics use \(255\) two-hot bins, with entropy coefficient \(\xi=3\times10^{-4}\). Finally, the PID gains are \((\alpha,\!\kappa,\!\tau)\!=\!(1,\!0.2,\!0.5)\), with \(\lambda^{\mathsf{max}}\!=\!1\) and \(b\!=\!0\). 
\begin{figure*}[t!]
\vspace{-25 pt}
    \begin{subfigure}{0.21\textwidth}
        \includegraphics[width=\linewidth]{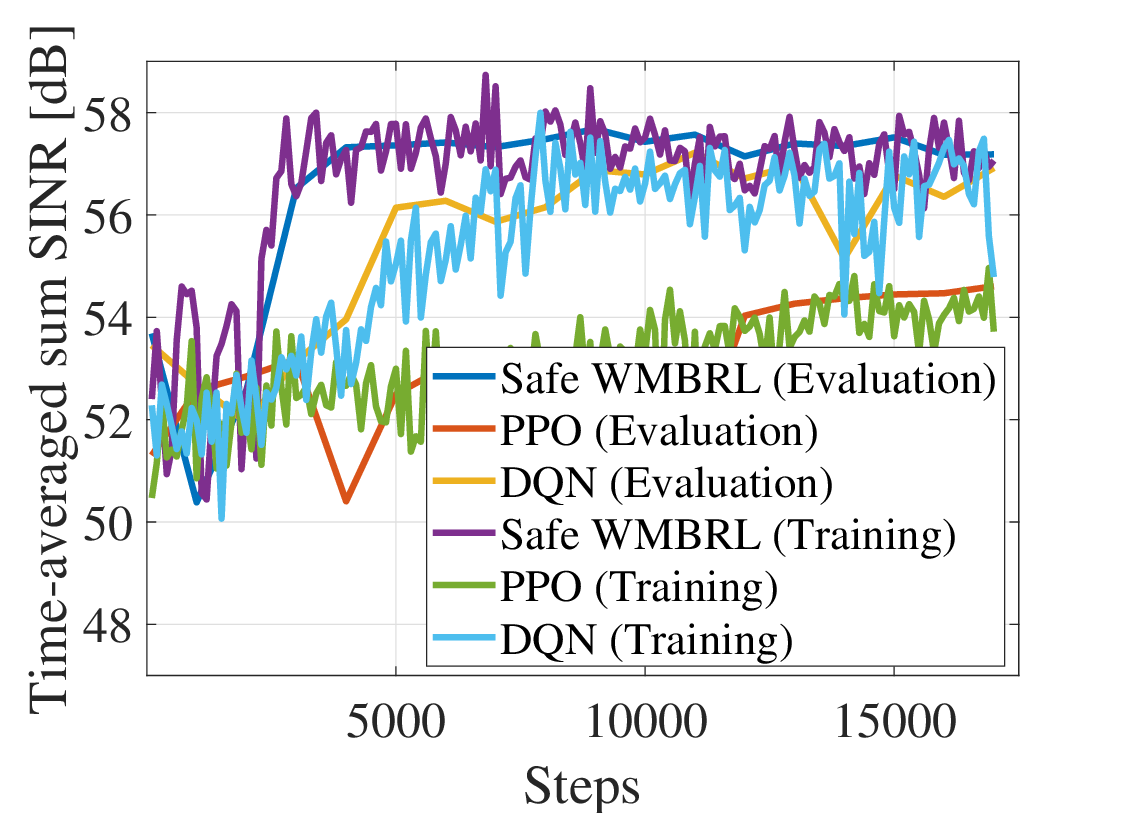}
        \caption{}\label{2a}
    \end{subfigure}
    \hspace{-15pt}
    \begin{subfigure}{0.22\textwidth}
       \includegraphics[width=\linewidth]{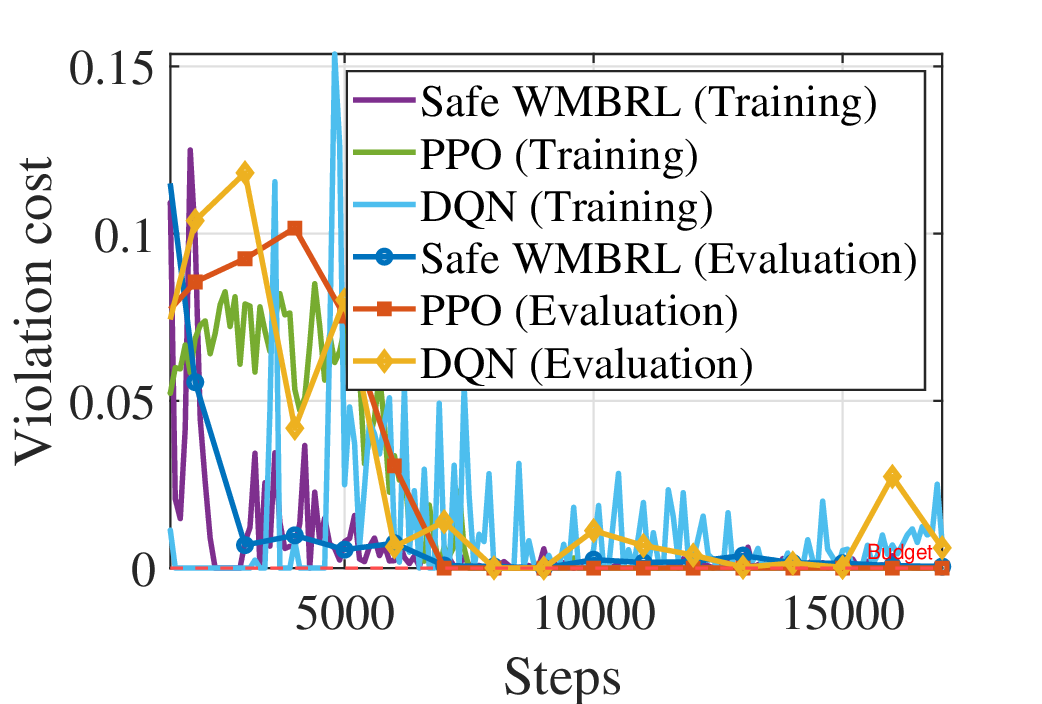}
        \caption{}\label{2b}
    \end{subfigure}
    \hspace{-16pt}
    \begin{subfigure}{0.22\textwidth}
        \includegraphics[width=\linewidth]{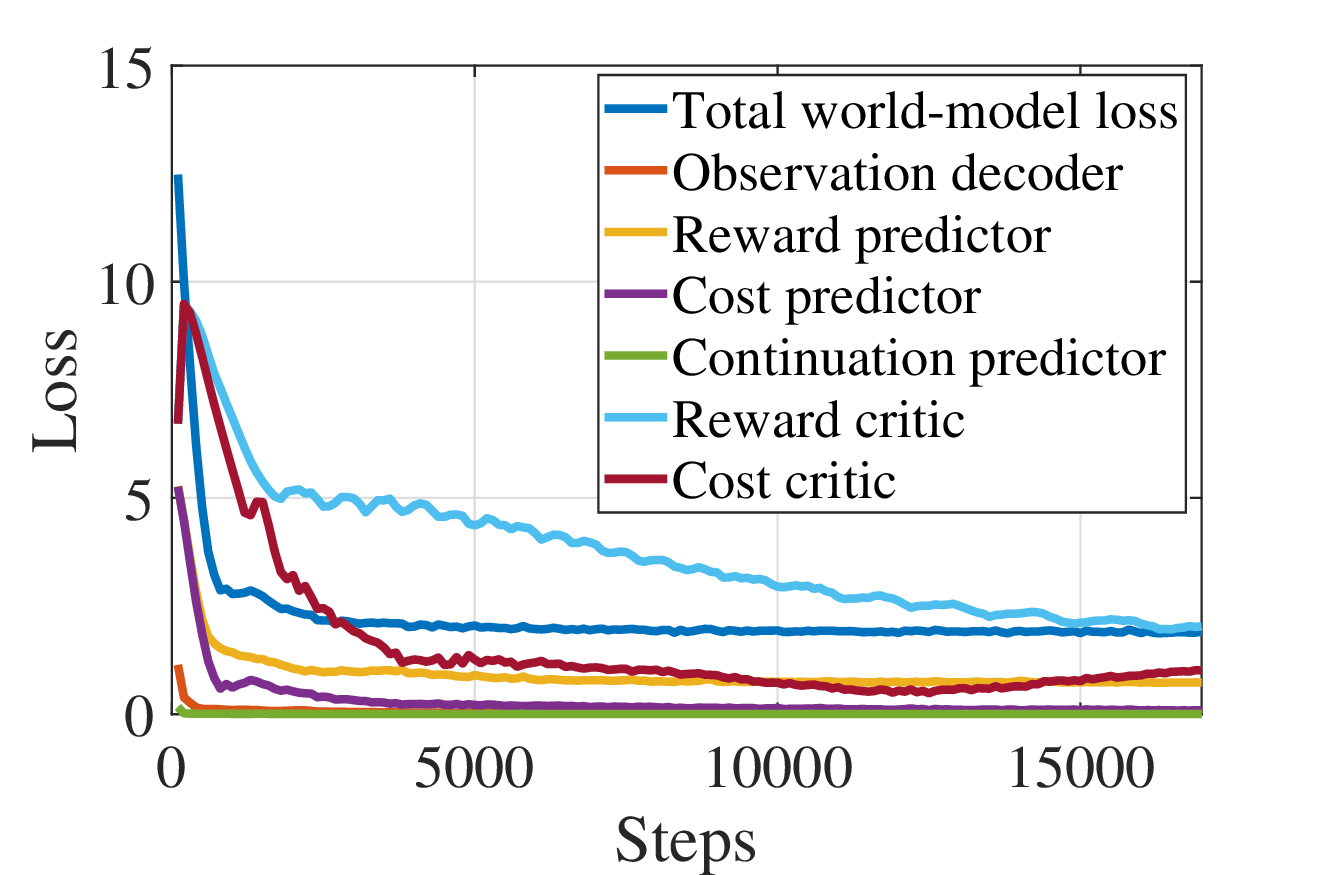}
        \caption{}\label{2c}
    \end{subfigure}
    \hspace{-15pt}
    \begin{subfigure}{0.23\textwidth}
        \includegraphics[width=\linewidth]{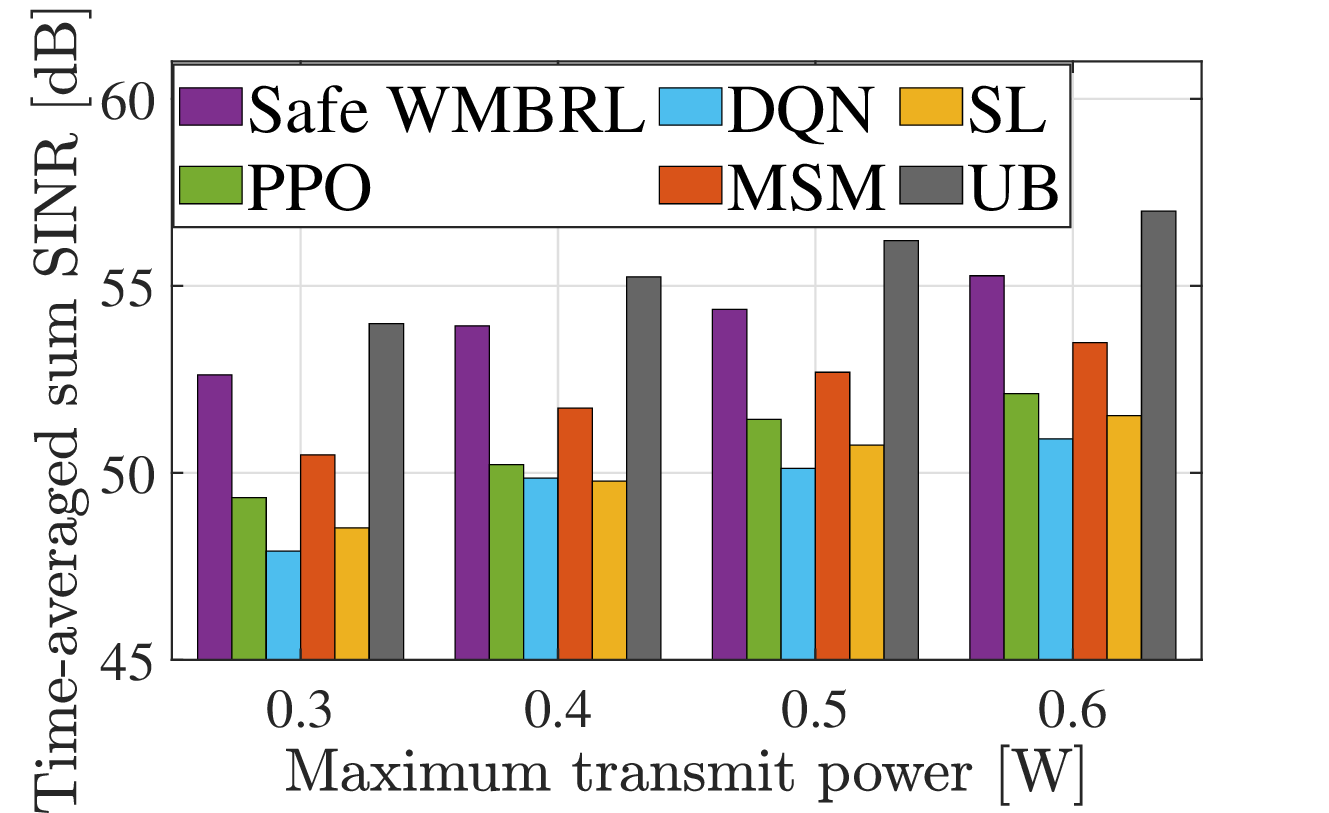}
        \caption{}\label{2e}
    \end{subfigure}
     \hspace{-16pt}
    \begin{subfigure}{0.20\textwidth}
        \includegraphics[width=\linewidth]{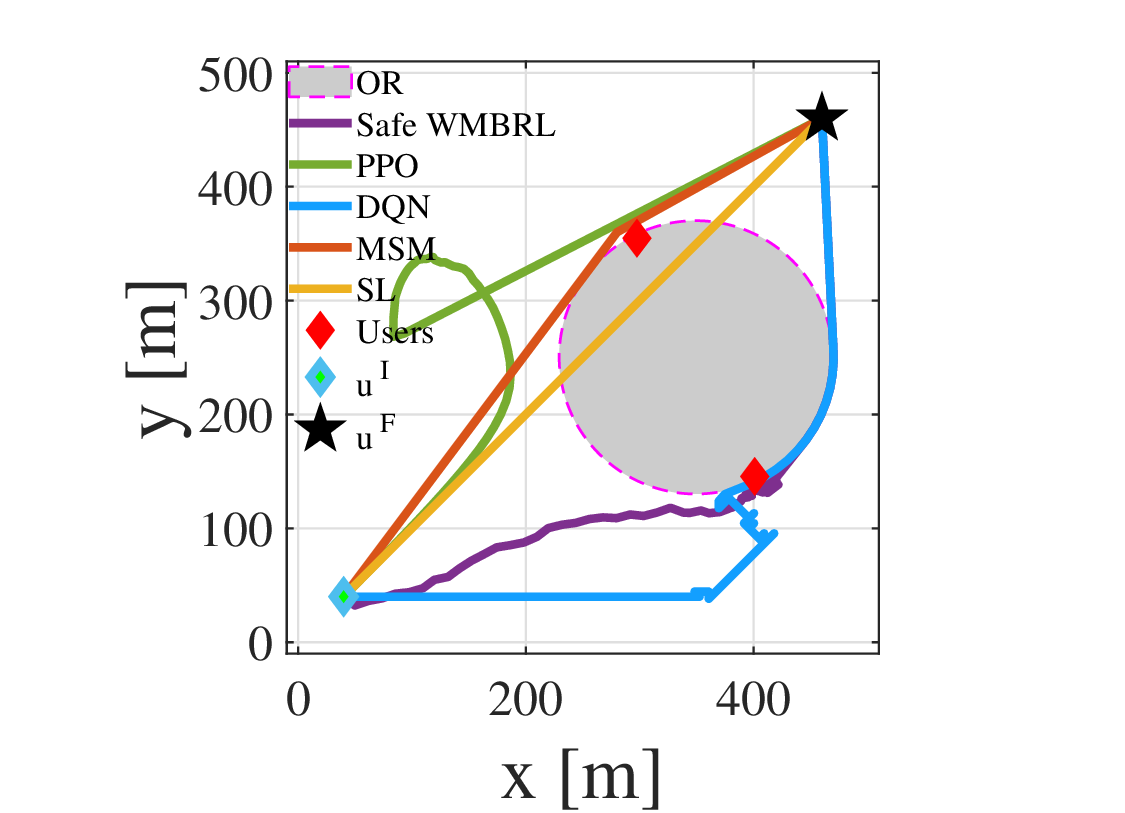}
        \caption{}\label{2d}
    \end{subfigure}
    \vspace{-19pt}
        \caption{\small Performance evaluations: (a) time-averaged sum SINR versus training steps, (b) violation cost versus training steps,  (c) learning losses versus training steps, (d) time-averaged sum SINR versus the maximum transmit power (\(P^{\mathsf{max}}\)), and (e) MBS trajectories.\vspace{-15 pt} \vspace{0 em}}
    \label{elem_var}
\end{figure*}

We evaluate the proposed WMBRL framework against four benchmark schemes: {\it PPO} and {\it deep Q-network (DQN)} with Lagrangian-based constraint handling, {\it move-stop-move (MSM)}, and {\it straight-line (SL)} trajectories. In the MSM scheme, the MBS moves to a feasible location with favorable channel conditions, remains stationary for a portion of the operation period, and then proceeds to its final position. The SL scheme follows a direct trajectory with uniformly spaced positions. In addition, an upper-bound (UB) benchmark is obtained by neglecting inter-user interference and assuming that each user is served directly from overhead with the full transmit power.

Fig. \ref{2a} shows that the proposed WMBRL framework achieves higher time-averaged sum SINR and faster convergence than PPO and DQN by optimizing policies over imagined trajectories, reducing the need for direct environment interactions. In contrast, DQN requires more environment interactions to converge, while PPO converges more slowly to a lower SINR. Fig. \ref{2b} shows that WMBRL rapidly reduces the violation cost to nearly zero and keeps it near zero, whereas DQN exhibits greater fluctuations and PPO converges more slowly. Furthermore, Fig. \ref{2c} shows that the total world-model loss decreases rapidly and stabilizes during training, while the decoder, reward, cost, and continuation prediction losses exhibit similar trends, indicating improved prediction accuracy. In addition, the reward- and cost-critic losses generally decrease during training, indicating improved estimation of long-term rewards and safety costs.

Fig. \ref{2e} shows that the time-averaged sum SINR increases with \(P^{\mathsf{max}}\) for all schemes. The proposed WMBRL framework consistently outperforms PPO, DQN, MSM, and SL, while achieving performance close to the upper bound.  Finally, Fig. \ref{2d} illustrates that WMBRL follows an adaptive trajectory that avoids the OR and reaches the prescribed final position. In contrast, DQN follows a less efficient trajectory with unnecessary deviations. Moreover, PPO, DQN, and MSM all avoid the OR, whereas the SL trajectory intersects it and thus violates the avoidance constraint.
\vspace{-5 pt}
\section{Conclusion}
In this paper, we investigated the joint optimization of MBS trajectories and digital beamforming to maximize the time-averaged sum SINR under mobility, OR avoidance, and transmit-power constraints. We developed a SafeDreamer-based WMBRL framework that learns system dynamics and optimizes the control policy through imagined trajectories while handling constraints using feasibility-preserving action mappings and safety-cost signals. Simulation results demonstrated that the proposed framework outperforms the considered benchmarks in terms of time-averaged sum SINR, while achieving higher sample efficiency and more reliable constraint satisfaction than the MFRL baselines.
\vspace{-5pt}
\bibliographystyle{IEEEtran}
\bibliography{bib.bib}
\end{document}